\documentclass[journal]{IEEEtran}
\usepackage{algorithm}
\usepackage{algorithmic}

\usepackage{amssymb}
\usepackage{amsmath}
\usepackage[makeroom]{cancel}
\usepackage{graphicx}
\usepackage{enumerate}
\usepackage{verbatim}
\usepackage{url}
\urldef{\mailsa}\path|{gurgurka}@boun.edu.tr|
\usepackage{siunitx}
\usepackage{epsfig}
\usepackage{caption}
\usepackage{subcaption}
\usepackage{mathtools}
\usepackage{color,soul}
\usepackage{mwe}
\usepackage{alltt}

\usepackage{listings}
\usepackage{graphicx}
\usepackage{caption}
\usepackage{subcaption}
\usepackage{tikz}
\usetikzlibrary{arrows,automata,positioning}
\usetikzlibrary{shapes}

\newcommand*\mycircledblue[1]{\tikz[baseline=(char.base)]{
            \node[shape=circle,draw,fill={rgb,255:red,37; green,128; blue,173}, text=white,inner sep=1pt] (char) {#1};}}

\newcommand{\lambdaNHU}{\lambda_{N_{HU}}}

\newcommand{\BSUnivPower}{P_{BS(U)}}
\newcommand{\DTwoDPower}{P_{D2D}}
\newcommand{\localPower}{P_{loc}}

\newcommand{\maxcontent}{N_c}

\newcommand{\be}{\begin{equation}}
\newcommand{\ee}{\end{equation}}
\newcommand{\bea}{\begin{eqnarray}}
\newcommand{\eea}{\end{eqnarray}}
\newcommand{\ba}{\begin{array}}
\newcommand{\ea}{\end{array}}
\newcommand{\bt}{\begin{tabular}}
\newcommand{\et}{\end{tabular}}
\newcommand{\bfin}{\begin{figure}}
\newcommand{\bfi}{\begin{figure}[!htb]}
\newcommand{\efi}{\end{figure}}

\newcommand{\BSPower}{P_{BS}}

\newcommand{\size}[1]{s_{(#1)}}

\newcommand{\profitAll}[1]{E_{all}^{(#1)}}
\newcommand{\profitLocal}[1]{E_{loc}^{(#1)}}
\newcommand{\profitDTwoD}[1]{E_{D2D}^{(#1)}}
\newcommand{\profitBS}[1]{E_{BS}^{(#1)}}
\newcommand{\profitBSUniv}[1]{E_{BS(U)}^{(#1)}}

\newcommand{\weightLocal}[1]{p_{loc}^{(#1)}}
\newcommand{\weightDTwoD}[1]{p_{D2D}^{(#1)}}
\newcommand{\weightBS}[1]{p_{BS}^{(#1)}}

\newcommand{\probI}{p_i}
\newcommand{\probJ}{p_j}
\newcommand{\probK}{p_k}

\newcommand{\fitfunctionDevice}{F_{loc}^{fit}}
\newcommand{\fitfunctionBS}{F_{BS}^{fit}}

\newcommand{\CapacityLocal}{C_{loc}}
\newcommand{\CapacityDTwoD}{C_{D2D}}
\newcommand{\CapacityBS}{C_{BS}}
\newcommand{\CapacityBSUniv}{C_{BS(U)}}

\newcommand{\RDTwoD}{R_{D2D}}

\newcommand{\HUDeviceCacheSize}{\mathbf{C}_{\mathbf{Dev}}}
\newcommand{\BSCacheSize}{\mathbf{C}_{\mathbf{BS}}}

\newcommand{\Nneighbours}{N_{ngh}}

\newcommand{\deltaDev}{\delta_{Dev}}
\newcommand{\deltaBS}{\delta_{BS}}
\newcommand{\PDC}{PDC}

\begin{document}

\title{Energy Prioritized Caching for Cellular D2D Networks}

\author{\IEEEauthorblockN{S. Sinem Kaf{\i}lo\u{g}lu, G\"{u}rkan G\"{u}r$^\dagger$ and
Fatih Alag\"{o}z}\\
\IEEEauthorblockA{Department of Computer Engineering, Bogazici University\\ Istanbul, Turkey\\
$^\dagger$Zurich University of Applied Sciences (ZHAW)\\ Winterthur, Switzerland\\ Email: \{sinem.kafiloglu, fatih.alagoz\}@boun.edu.tr}, gueu@zhaw.ch
\thanks{This work was supported by the Scientific and Technical Research Council of Turkey (TUBITAK) under grant number 116E245.}
}

\maketitle
\begin{abstract}
Multimedia content transmission heavily taxes network resources and puts a significant burden on wireless systems in terms of capacity and energy consumption. In this context, device-to-device (D2D) paradigm has a utilitarian value to alleviate the network burden by utilizing short-range transmissions with less energy cost. For the realization of proper D2D networking, caching in devices is essential. Additionally, caching needs to operate efficiently with the aim of realizing network-wide energy improvements. To this end, we study multimedia caching in cellular D2D networks and propose an energy prioritized D2D caching ($EPDC$) algorithm in this work. We also investigate the optimal caching policy in that system setting. The impact of device capacity and D2D transmission range on energy consumption is studied by focusing on different operation modes such as D2D and base station transmissions. According to the simulation results, $EPDC$ algorithm has substantial energy-efficiency gains over the commonly utilized Least Recently Used ($LRU$) algorithm and content-based caching  algorithms $\PDC$ and $SXO$. For larger D2D transmission ranges, this improvement becomes more evident.
\end{abstract}


\section{Introduction}

The multimedia traffic in data networks has surged with the exploding video consumption of users and substantially enriched services of multimedia providers. Based on the Cisco Annual Internet Report forecasts, multimedia continues to be of enormous demand today, and there will be escalated burden with the application requirements of the future with 8K and VR streaming, and cloud gaming even beyond the forecast period of 2023 in 5G networks~\cite{CISCOFORECAST}. In this context, the network operators have to meet certain QoS levels for provisioning consumer-driven content-based services. Nevertheless, the network resources are limited and burgeoning content consumption requests are observed leading to spiking energy consumption. In that regard, the energy efficiency (EE) problem emerges in these content-flooded networks.  This problem needs to be resolved by practical networking paradigms in an efficient manner.

In-network caching is one of the techniques to alleviate the EE problem in future wireless networks. In this approach, the network load is reduced while transmission latency is lowered~\cite{6364320}. In~\cite{INNETWORKCACHE}, an energy-efficiency based in-network caching algorithm is proposed. It can perform better than the existing algorithms in term of both energy consumption and cache hit rate according to their simulation results.  
Fang et. al. also investigated in-network caching for achieving green content centric network paradigm~\cite{ENERGYINNETWORK}. 
Overall, in-network caching is a promising solution to tackle the EE challenges. 
In the same vein, Device-to-Device (D2D) communication technique is beneficial in terms of various aspects such as delay reduction, improved data rates, boosted spectral efficiency in addition to reduced energy consumption~\cite{PRERNA202048}. D2D mechanism can be used as a relaying factor to improve the power-efficiency of cellular networks~\cite{doi:10.1002/ett.3576}. It also improves the radio resource utilization to alleviate the communication coverage challenges in cellular systems~\cite{ZHANG201782}. Therefore, it is posed as a crucial enabler for boosting system capacity for heavy multimedia workloads and improving network-wide energy efficiency~\cite{7904715}.

The simultaneous exploitation of edge caching and D2D networking is broadly studied, e.g. ~\cite{PIMRC}. The major part of the literature focuses on the service improvement as a crucial element in cache-aided D2D networks~\cite{throughputD2D,successprobD2D}. In that regard, some studied metrics are throughput~\cite{throughputD2D}, total service success probability~\cite{successprobD2D}, transmission delay~\cite{delayandoffload} and offloading rate~\cite{delayandoffload,offloadingandsumrateD2D}. 
Furthermore, the energy consumption and efficiency pose as vital metrics to be examined in cache-enabled D2D networks. In~\cite{7904715}, the energy efficiency in a D2D caching HetNet is investigated in great detail without focusing on  D2D caching optimization. 
On the contrary, in~\cite{GUR201533} a greedy caching algorithm is proposed and compared to the optimal case in terms of energy efficiency. That latter work resides not on D2D networks but Information-Centric Networks (ICNs). 
Lee et. al. propose caching policy and cooperation distance design in D2D caching networks for \textit{i)} energy efficiency \textit{ii)} throughput optimization separately in~\cite{EE_OPT_CACHED2D}. However, their energy model is limited to the device reception powers only.

In this work, we investigate the edge cache management techniques in D2D cellular networks owing to the fact that in-network caching is regarded as an efficient facilitator for improving system-wise energy consumption. We have a comprehensive service-mode based energy model and define content energy expenditures based on the prospective consumption across the network. We formulate the optimization for the cache replacement problem in D2D networks and solve it with dynamic programming. Besides, we propose a heuristic algorithm \textit{Energy Prioritized D2D Caching (EPDC)} to improve the time complexity of the content services. Finally, a rigorous performance analysis is performed while focusing on the service rate and energy consumption in different operation modes separately.

The key contributions of this work are as follows:
\begin{itemize}
	\item We introduce a new energy expenditure model based on prospective energy consumption for content retrieval in cellular D2D networks. 
	\item We optimally manage caches for the replacement problem in our D2D network architecture with the aim of keeping contents that would induce large energy cost.  
	\item We propose an energy cost based heuristic, namely \textit{Energy Prioritized D2D Caching (EPDC)} algorithm. 
	\item We investigate the impact of cache capacity and D2D transmission range on the service capacity and energy consumption of several operation modes. 
\end{itemize}

\section{Related Work}
There is a plethora of work in the literature on \textit{content caching in D2D networks}. We start our investigation by elaborating on papers for content-based D2D cache management.  
In~\cite{article}, they have shown that Mandelbrot-Zipf (MZipf) distribution scale law in cache-aided D2D networks is the same with commonly utilized Zipf distribution and hence already existing studies utilizing the Zipf distribution become practically usable. Besides, they have shown that D2D cache network has improvement over the alternative BS unicast scenario. Congruently, we make use of the Zipf distribution in our content popularity model. However, instead of a comparison with no cache-aided D2D scenario, we specifically focus on the cache replacement for the energy aspect and make our comparison to conventional cache replacement techniques. Chen et. al. learn user preferences and manage caching accordingly, and they achieve a remarkable caching gain over the baseline that utilizes content popularity~\cite{8425746}. They exhaustively study the offloading probability while we focus on the energy consumption in great detail and inspect consumption in modes separately. Li et. al. propose caching in Fog Radio Access Network (f-RAN) enhanced with D2D technology~\cite{8746278}. Based on social preferences, AP communities are constructed and within a community, cooperative caching is utilized while communities are dynamically re-determined. For further improvement in transmission delay, D2D mode transmission is used and most appropriate device in every AP and most requested contents are selected for the caching scheme. They interrogate cache hit and delay metrics while in our study we deliberately focus on the network-wise energy consumption. Besides, we have a more general network architecture that can be tuned for custom network scenarios such as fog architecture if needed. 
Overall, these studies are on the content-based caching in D2D networks.

Now, we specifically discuss \textit{energy} focused caching studies. The greening of computer networks becomes more challenging with the advent of 5G networking due to high multimedia traffic and low latency requirements~\cite{8014295}. To overcome the energy burden of 5G heterogeneous networks, in-network caching is broadly utilized in different segments. Yang et. al. propose a self-optimizing in-network caching algorithm for improving energy-efficiency in 5G networks~\cite{doi:10.1002/ett.3221}. However, the D2D communication mechanism is not utilized in their study. As D2D paradigm is a major component in our system, we particularly focus on the D2D network caching studies. Chen et. al. propose to place contents via mobility-awareness to optimize cache hit performance in~\cite{8067654}. Apart from that, they optimize content transmission powers of small base stations (SBSs) and devices to alleviate the energy cost. They demonstrate the improvement in terms of cache hit and energy efficiency in contrast to random, popular and mobility-aware caching with fixed data amount (MCF). Similar to our study, they interrogate the general 5G environment from the perspective of caching and EE. However, we restrain ourselves to a more specific 5G domain that is the  ``edge D2D network". Above all, they analyze the cache placement problem while we focus on the replacement. 
In~\cite{8108248}, they define an energy ratio metric and propose a sub-optimal caching scheme built upon this metric to improve EE. According to their study, their proposal performs better than the equal probability random, most popular random and cut-off random caching techniques in terms of EE. They have restricting assumptions such as no energy consumption in cache hits and the lack of universal source concept. Besides, we define EE as the total energy consumption over the successfully received data while 
they measure differently as the ratio of the energy in caching network over no-caching network.

Finally, we elaborate on some optimization studies in D2D caching networks. Chen et. al. optimize the caching and resource allocation for maximizing the offloading of services to local hits or from neighbouring devices in D2D operation mode~\cite{D2D_ZIPF2}. However, this study is limited to offloading probability inspection only whereas we focus both on the service capacity and energy consumption of our entire D2D cellular network. 
In~\cite{8651586}, they study cache partitioning, content placement and user association problems in D2D HetNets. First, they propose a game-based cache partitioning scheme to achieve the optimal cache space allocation. Building upon this, they formulate a service delay minimization problem for content placement and user association. Due to the mixed integer nonlinear nature of this problem, Lagrangian partial relaxation and partitioning into sub-problems is applied and solved accordingly. 
Even though this is a holistic caching optimization study in D2D networks, it lacks the energy efficiency research that is a very fundamental premise of our work. Besides, our content model is more realistic with chunk and layer dimensions that are not available in~\cite{8651586}. 
Choi et. al. propose a cache placement algorithm to maximize the overall content quality observed by users and also a resource allocation algorithm with the objective of maximizing the expected requested content quality of users constrained by tolerable delay~\cite{8374957}.
They consider different content qualities in D2D caching networks and focus on the content quality while our main objective is to improve the energy efficiency of D2D network. In~\cite{7931641}, Chen et. al. propose optimal proactive caching for improving offloading to D2D and also optimize D2D transmission power to maximize service percentage. Even though they consider overlaying in D2D communications and investigate the D2D offloading gain and device energy consumption, their content model is limited. We utilize layering and chunking in our content model yielding a more comprehensive content structure. 

In a nutshell, the core motivation of our study is to optimize the cache replacement problem of popularity, partition and quality based multi-attributed multimedia contents for minimizing the energy burden of D2D edge networks.

\section{System Model}
\label{sec:networkSystem}
\begin{figure}[ht]
\centering
\includegraphics[width=0.9\columnwidth]{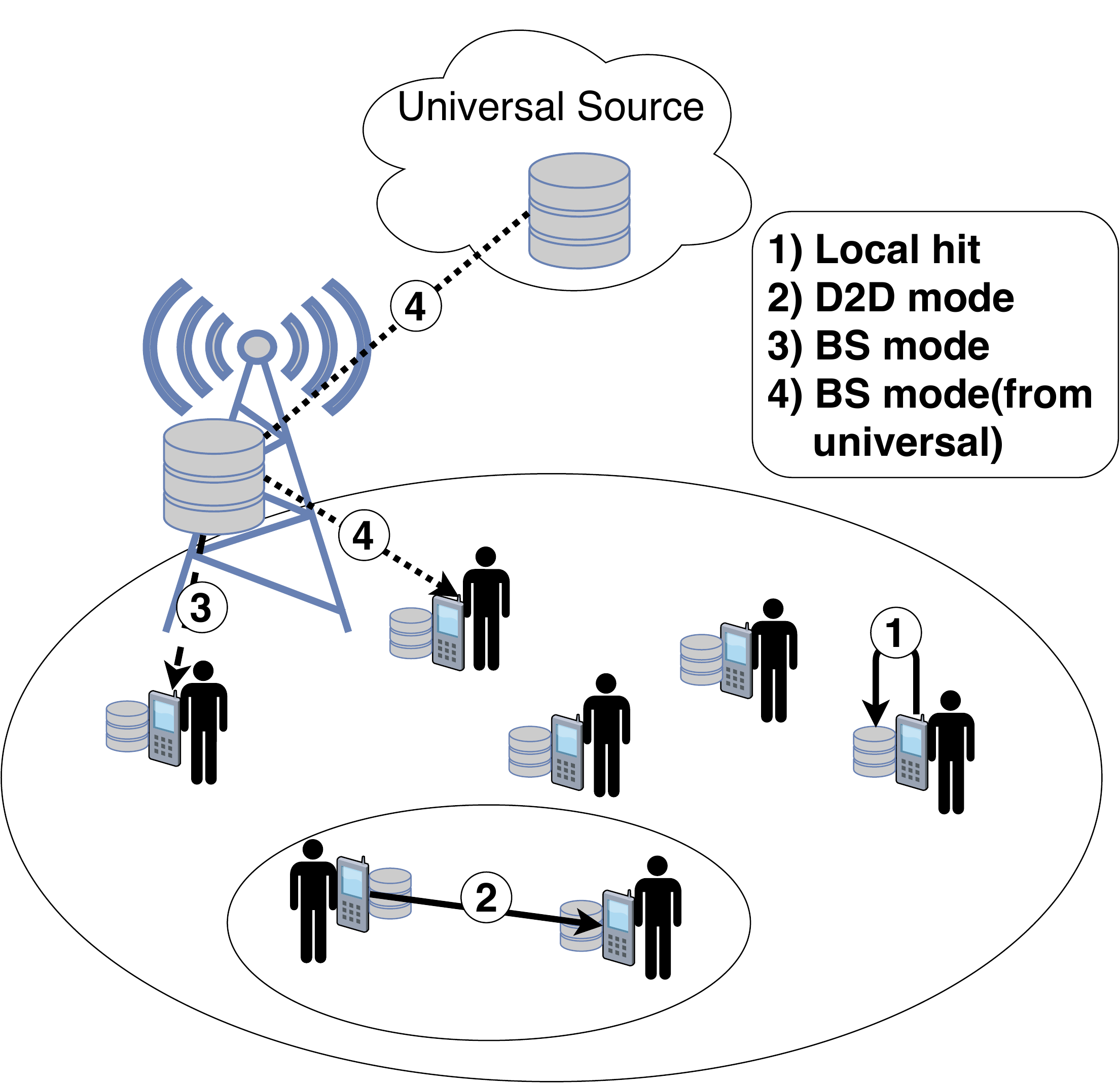}
\caption{Network architecture.}
\label{fig:sysModel}
\end{figure}

In this section we present the network architecture, content model and content request management in our system. 
The network architecture is shown in Fig.~\ref{fig:sysModel}. In our network architecture, there is one cell with one base station (BS) and D2D enabled user devices are scattered across this cell. As broadly utilized in the literature~\cite{PPP,PPP2}, devices are spread across the cell according to Point Poisson Process (PPP) with mean density $\lambdaNHU$. There exists the universal source to serve the contents that are not available in any system unit (e.g. BS, device) at the edge network. In our network architecture, there are four main service operation modes: \mycircledblue{1} local hit, \mycircledblue{2} D2D mode, \mycircledblue{3} BS mode (direct), \mycircledblue{4} BS mode (from the universal source) as marked in Fig.~\ref{fig:sysModel}.

In our content model, request probabilities are determined by popularity that is based on the Zipf distribution. The probability of content requests are calculated via $Zipf(s,\maxcontent)$ where $s$ determines the distribution skewness and $\maxcontent$ is the total number of contents in the system. 
Content chunking is also used in communication systems with the aim of performance improvement~\cite{opt_chunking,chunking2}. Accordingly, we utilize chunked contents in our model. Inter-chunk popularity differentiation is another approach used for improving transmission and caching performance in content-driven networks. In that regard, the request rate of content chunks are calculated by 
the commonly used $Weibull(\lambda,k)$ distribution with the scale and shape parameter $\lambda$ and $k$ respectively.

Our contents are scalable coded (SVC) videos. Each content is of high quality (HQ) consisting of one \textit{base} and one \textit{enhancement} layer. For standard quality content consumption, the base layer is sufficient whereas both layers are required for HQ visualization. The request rate for HQ contents in our network scenario is $p_k \in [0,1]$. 
We define each layer of a given content's particular chunk uniquely as a \textit{content unit tuple} by its content, chunk and layer ids respectively as $\{i,j,k\}$. Each such tuple is mapped to a unique \textit{content unit identifier} $u$ for tractability. 
The average content sizes are 322 and 474 Mbits for SQ and HQ SVC videos, respectively~\cite{PIMRC}. For the calculation of these values, QCIF-formatted and an hour long temporal scalable encoded multimedia are exploited~\cite{5342293}. In~\cite{reisslein}, the calculation formula for the mean video frame sizes of different layers is provided. Some utilized video samples are $Citizen~ Kane$, $Jurassic~ Park~ I$, $Star~ Wars~ IV$, $Aladdin$ and $Tonight~ Show$. For the full video list, please refer to our previous work~\cite{PIMRC}. 
In Section~\ref{sec:alg}, we preserve the content characteristics of our content model and only change the total number of contents to push the system to its limits for the feasibility inspection of the time complexity analysis. In the performance evaluation section, we also keep the total number of contents fixed and investigate the impact of some other system parameters as described in Section~\ref{sec:results}.

\begin{figure}
	\centering
	\includegraphics[width=0.9\columnwidth]{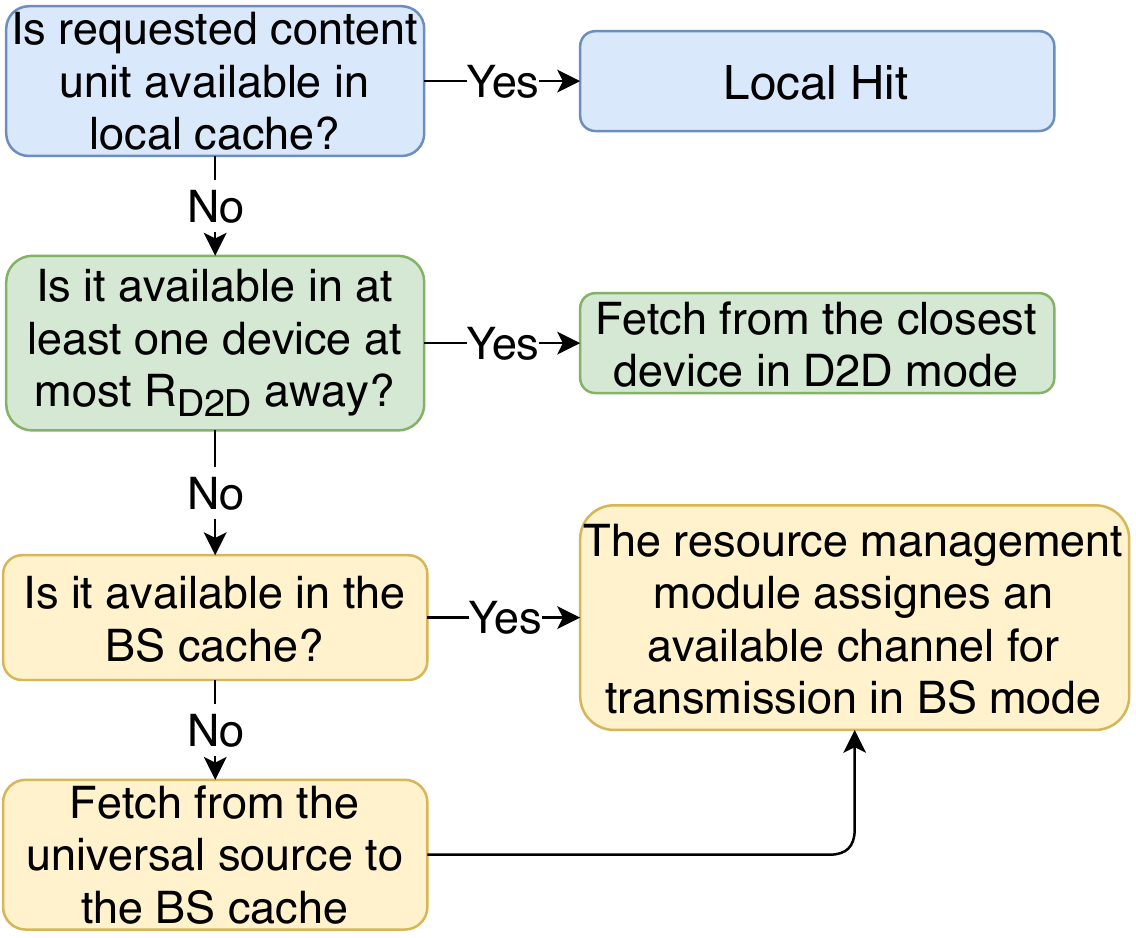}
	\caption{Content request management in the edge network.}
	\label{fig:requestManagement}
\end{figure}

We extend the wireless D2D network in our previous work~\cite{PIMRC} with cellular support. 
In this context, the BS transmission power $\BSPower$ and BS transmission path loss exponent $n_{\theta_{BS}}$ are introduced for the calculation of the BS power strength in the BS channel model. The general content request management scheme is shown in Fig.~\ref{fig:requestManagement}. First, the local cache is looked up for a requested content unit. If that unit is not found in the local cache, then D2D mode is chosen as long as that requested unit is available in at least one device in the reception range of (at most $\RDTwoD$ away from) the requester device. We assume the inter-device distances are known by user devices according to a signaling scheme to discover neighbour devices~\cite{8403950}. Thereby, devices in the reception range of each other can 
determine inter-node distances.
The BS serves when that requested unit is not found in any device in the reception range of the requester. This case branches into two scenarios: \textit{i)} that unit is available in the BS cache, \textit{ii)} that unit is not available in the BS cache. In the scenario-\textit{i}, the BS cache directly serves that unit. In the scenario-\textit{ii}, that unit is first fetched from the universal source to the BS and then served to the requester to allow access to the exterior of the edge network system. Note that the multi-mode service operations (local hit, D2D mode, BS mode) in cellular D2D networks are applicable as shown in~\cite{SINEMKAFILOGLU2020107083}. 
All requests and network using operation modes (\mycircledblue{2}-\mycircledblue{4}) acquire the same network access priority. If a content unit cannot be served due to all channels being busy, then that unit and prospective units of that content are all dropped.

The BS and devices have storages for caching multimedia contents. 
In this paper, we mainly study on the cache replacement techniques running both on user devices and the BS.
In the cache replacement procedure, the system unit (device/ BS) evicts its cache until there is sufficient free space for the requested and retrieved video content unit. The decision for the evictions will be explained in the next section.

\section{Optimal and Energy Prioritized D2D Caching (EPDC)}
\label{sec:alg}

5G systems is posed to be an enabler for connected multimedia services. However, the tremendous demand for the multimedia services raises the energy consumption of the wireless networks, leading to a fundamental cost and environmental issue that needs to be addressed. To this end, 5G communications need to improve energy efficiency performance~\cite{7414384}. The D2D edge network is one the 5G technologies that provides service capabilities closer to end users and thus reduces the energy burden on the network. 
Similarly, in-network caching is an important mechanism to reduce network transmissions and enhance the energy efficiency as well. 
In that regard, we formulate our cache replacement problem in D2D edge network for decreasing energy consumption in this section. Subsequently, optimal and heuristic approaches are employed to solve it for our system setup.

Initially, we solve the scalable video caching in D2D networks with the backhaul support optimally. For the ease of tractability, the model and system parameters are listed in Table~\ref{table:Notations}.  Say the content units residing in the local cache of some user are denoted by $C$. Some user requests a content unit $u'$ and it is retrieved from some other system component via wireless transmission. If the size of content units in the set $C$ and new arrived unit $u'$ together exceed the cache capacity $\HUDeviceCacheSize$, then a subset of content units $C$ residing in the local cache need to be replaced. The eviction decision is essentially a 0/1 Knapsack problem~\cite{GUR201533}. For each content unit you either decide to sustain that unit in the local cache or remove it. The content units that are decided to remain in the local cache are designated by $\overline{C}$. In our network architecture at each caching decision phase, we want to minimize prospective energy consumption. Thereof, we sustain content unit set $\overline{C}$ that would induce the largest total energy expenditure. This way, content units with lower energy consumption will be evicted. In the prospective requests for them, they will be fetched from other system components with less cost. 
The optimization problem is stated as follows:
\begin{align}
& \underset{}{max} \sum_{u \in C} \profitAll{u}~1_{[u \in \overline{C}]} \label{eq:objective}\\
&  ~~ s.t.~~ \size{u'} + \sum_{u \in C} \size{u}  ~1_{[u \in \overline{C}]} \leq \HUDeviceCacheSize \label{eq:constraint}
\end{align}
\begin{table}[ht]
\centering
	\caption{Notations.}
	\begin{tabular}{ | p{1.2cm} | p{6.6cm} | }
			\hline
			\textbf{Parameter} & \textbf{Explanation} \\ \hline
			$\maxcontent$ & The total number of contents \\ \hline
			$s$ & The Zipf distribution skewness parameter \\ \hline
			$\lambda$ & The Weibull distribution scale parameter \\ \hline
			$k$ & The Weibull distribution shape parameter \\ \hline
			$p_i$ & The probability of content $i$ being requested \\ \hline
			$p_j$ & The probability of content chunk $j$ being requested  \\ \hline
			$p_k$ & The probability of high quality contents being requested  \\ \hline
			$\{i,j,k\}$ & The unique \textit{content unit tuple} (The $i^{th}$ content's chunk $j$ of layer $k$) \\ \hline
			$u$ & The unique \textit{content unit identifier} such that each $\{i,j,k\} \mapsto u$\\ \hline
			$s_{(u)}$ & The size of the content unit $u$ \\ \hline
			$\RDTwoD$ & The maximum allowable distance between a requester and transmitter for D2D operation\\ \hline
			$\Nneighbours$ & The number of neighbouring devices at most $\RDTwoD$ away from a requester \\ \hline
			$\HUDeviceCacheSize$ & The device cache capacity  \\ \hline
			$\BSCacheSize$ & The base station cache capacity  \\ \hline
			$\CapacityLocal$ & The expected service capacity of content unit local hit \\ \hline
			$\CapacityDTwoD$ & The expected service capacity of content unit retrievals via D2D technique  \\ \hline
			$\CapacityBS$ & The expected service capacity of content unit retrievals from the BS cache  \\ \hline
			$\CapacityBSUniv$ & The expected service capacity of content unit retrievals from the universal source to the BS cache 
			\\ \hline
			$\localPower$ & The local hit service power level\\ \hline
			$P_{D2D}$ & The transmission power level of any device\\ \hline
			$\BSPower$ & The base station transmission power level \\ \hline
			$\BSUnivPower$ & The base station reception power level\\ \hline
			$\theta_{loc}$  & The local hit service power parameter\\ \hline
			$\theta_{BS}$  & The base station reception power parameter\\ \hline
			$\deltaDev$ & The incremental units for the device cache \\ \hline
			$\deltaBS$  & The incremental units for the BS cache\\ \hline
	\end{tabular}
	\label{table:Notations}
\end{table}
with the indicator function $1_{[x]}$ set to one if $x$ is true and zero otherwise. Our objective function $\sum_{u \in C} \profitAll{u}~1_{[u \in \overline{C}]}$ in (\ref{eq:objective}) gives the total energy consumption of units across the network that are not evicted from the cache set $C$ and residing in $\overline{C}$. With constraint (\ref{eq:constraint}), we ensure that the aggregate size of the new unit $u'$ and the units not evicted from the cache 
does not exceed the cache capacity. 
We formulate the prospective energy consumption of any content unit $u$ as $\profitAll{u}$
in~(\ref{eq:profit_all}). This energy is branched into the four scenarios that are shown in Fig.~\ref{fig:sysModel}: \mycircledblue{1} Local hit, \mycircledblue{2} Transmission via D2D technique, \mycircledblue{3} Transmission from the BS cache, \mycircledblue{4} Transmission from the universal source across the BS 

With the probability $\weightLocal{u}$ the first scenario occurs and 
$\profitLocal{u}$ will be the corresponding prospective energy expenditure for the local hit of content unit $u$. The second scenario consumes $\profitDTwoD{u}$ energy when the content unit will not be available in the local cache but retrievable from some other device with probability $(1-\weightLocal{u})\weightDTwoD{u}$. If no device in the transmission range (centered at the requester with radius $\RDTwoD$) can serve due to the lack of unit, the BS takes action and will serve the content unit from the BS cache with the energy expenditure level $\profitBS{u}$. 
Unless the relevant unit is in the BS cache, the transmission path from the universal source to the BS cache and then from there to the requester device will be utilized with $(1-\weightLocal{u}) (1-\weightDTwoD{u}) (1-\weightBS{u})$ probability and $\profitBSUniv{u}$ energy will be consumed. The 
expected energy consumption of unit $u$ based on listed four scenarios is given in~(\ref{eq:profit_all}). 
The functions corresponding to the probability of content unit availability for different system components are provided in the Appendix. 
\begin{align}\label{eq:profit_all}
       \profitAll{u}    &:= \weightLocal{u} \profitLocal{u}+ (1-\weightLocal{u}) \weightDTwoD{u} \profitDTwoD{u}\nonumber \\ &+ (1-\weightLocal{u}) (1-\weightDTwoD{u}) \weightBS{u} \profitBS{u}\nonumber \\ &+ (1-\weightLocal{u}) (1-\weightDTwoD{u}) (1-\weightBS{u}) \profitBSUniv{u}
\end{align}

The expected energy expenditure of the given four scenarios are listed as follows:

\begin{align}
   \profitLocal{u}    &:=  \localPower \cdot \frac{\size{u}}{\CapacityLocal}\label{eq:profit_local} \\
    \profitDTwoD{u}    &:=   \DTwoDPower \cdot \frac{\size{u}}{\CapacityDTwoD}\label{eq:profit_D2D}  \\
    \profitBS{u}       &:=   \BSPower \cdot \frac{\size{u}}{\CapacityBS}\label{eq:profit_BS}  \\
    \profitBSUniv{u}   &:= [\BSUnivPower \cdot \frac{\size{u}}{\CapacityBSUniv}] + \profitBS{u}\label{eq:profit_BS_U}     
\end{align}

The expected energy expenditures are calculated by the product of power in some service type and its expected service duration for a given content unit. During local hits, a device consumes the least energy compared to other system components. We define this power $\localPower$ as $\frac{P_{D2D}}{\theta_{loc}}$ while its expected service duration is $\frac{\size{u}}{C_{loc}}$. The largest service rate among the system components is $C_{loc}$.
The product of $\localPower$ and $\frac{\size{u}}{C_{loc}}$ gives the expected local hit energy consumption $\profitLocal{u}$  for some content unit $u$ as shown in~(\ref{eq:profit_local}). 
The product of the device transmission power $P_{D2D}$ and its expected prospective service duration $\frac{\size{u}}{C_{D2D}}$ gives the expected D2D transmission energy $\profitDTwoD{u}$ (in~\ref{eq:profit_D2D}) for unit $u$. $C_{D2D}$ is calculated by taking the average distance $\frac{R_{D2D}}{2}$ for future D2D transmissions. 
The devices do not know the cache status of others, and for future predictions they tune to average values for the calculation of prospective energy figures. 
In~(\ref{eq:profit_BS}), the expected energy consumption for a transmission of some unit $u$ from the BS cache is calculated similar to the $\profitDTwoD{u}$.
When we consider the fourth scenario, the content unit retrieval basically branches into two parts: \textit{i)} from the universal source to the BS cache, \textit{ii)} from the BS cache to the requester device.
The retrieval to BS consumes $\BSUnivPower \cdot \frac{\size{u}}{\CapacityBSUniv}$ expected energy while the rest has $\profitBS{u}$ as shown in~(\ref{eq:profit_BS_U}). 
For the retrieval to the BS cache, its reception power is $\BSUnivPower$:=$\frac{\BSPower}{\theta_{BS}}$ and its expected prospective reception period is $\frac{\size{u}}{\CapacityBSUniv}$.

\begin{algorithm}\footnotesize
\caption{Dynamic Programming Caching Algorithm}\label{alg:dyn}
\begin{algorithmic}
\STATE{\textbf{OPT(}$\boldsymbol{C,u',\HUDeviceCacheSize}$)\{}
\STATE{}
\STATE{$Capacity$ $\gets$ $TotalSize(C)$;}\\
\IF{($Capacity+\size{u'} \leq \HUDeviceCacheSize$)}
    \STATE{return $~C\cup\{u'\};$ \%the set kept in the cache}
\ELSE
    \STATE{\%traverse all content units in the local device cache}
    \FORALL{($i = 1:1:|C|$)}
        \STATE{\%traverse weights with $\deltaDev$ (= 0.01 Mbs) incremental units}
        \FORALL{$j = 1:1:\left\lfloor\frac{Capacity-\size{u'}}{\deltaDev}\right\rfloor$}
            \STATE{}
            \IF{((i == 1) $\lor$ (j == 1))}
                \STATE{~$E_{cum}(i,j)$ $\gets$ 0; \%first row and column are set to zero}   
            \ELSIF{($\size{i}<j$)}
                \STATE{\%take maximum cumulative prospective energy consumption of in-cache units by either excluding $i^{th}$ unit or including it}
                \STATE{{$E_{cum}(i,j)\gets max\{E_{cum}(i-1,j),E_{all}^{(i)}+E_{cum}(i-1,j-\size{i})\}$};}
            \ELSE
                \STATE{{$E_{cum}(i,j) \gets E_{cum}(i-1,j)$;}}
            \ENDIF
        \ENDFOR
    \ENDFOR
    \STATE{}
    \STATE{\%start from the last row and column of $E_{cum}$ backtrack and mark units that will continue to reside in the local device cache as $\overline{C}$}
    \STATE{$\overline{C} \gets$ backtrack($E_{cum}$);}
    \STATE{return $\overline{C}\cup\{u'\}$;}    
\ENDIF
\STATE{\}}
\end{algorithmic}
\end{algorithm}

Dynamic programming algorithm is one of the techniques to solve the 0-1 knapsack problem optimally~\cite{knapsack_dyn}. In our optimal solution, we utilize dynamic programming given in Algorithm~\ref{alg:dyn}. In this algorithm, if the cache capacity $\HUDeviceCacheSize$ does not suffice for the residing content units set $C$ and the newly requested one $u'$, then a subset of $C$ will be evicted for the sake of newly arrived unit $u'$. Our main motivation is to keep content units in the local cache that would otherwise result in large prospective energy consumption. We consider the cache as the knapsack. Each content unit $u$ is an item that is weighted by its size $s_{(u)}$ and valued by its energy consumption $E_{all}^{(u)}$.  
In dynamic programming, we break the optimization problem into subproblems, recursively solve these subproblems and store their results in $E_{cum}$ array as given in Algorithm~\ref{alg:dyn}. By backtracking $E_{cum}$, we select the units with the largest energy sum to continue residing in the cache. These units are designated by the set $\overline{C}$.
However, the time complexity of this algorithm is $O(|C| \cdot W)$ where $W$ is the residual cache capacity after the new unit is inserted and $|C|$ is the number of content units in a given device cache~\cite{GUR201533}. This complexity does not provide a practical calculation time. Thereof, we propose a new energy prioritized D2D caching algorithm given in Algorithm~\ref{alg:EPDC}. 
The core idea of this greedy algorithm is to utilize our analytically calculated expected energy consumption of units $\profitAll{i}$ 
to alleviate the energy burden on the network system. 
 In this heuristic, at the eviction decision phase, all content units in the set $C$ are sorted based on their 
 $\profitAll{i}$ from the largest to the smallest one. 
 As shown in Algorithm~\ref{alg:EPDC} starting from the least energy consuming unit $c_k$, 
units are eliminated from the cache until the free space is suitable for storing new unit $u'$. Thus, unit(s) with low energy burden across the network are eliminated first. Large energy consuming units over the network are preserved locally for less energy burden in future requests. Therefore, this scheme is named as energy prioritized D2D caching (EPDC).
The time complexity of this procedure is bounded by the sorting procedure of elements in the set $C$ based on their $\profitAll{i}$ values. It is $O(|C| \cdot log |C|)$ with $|C|$ denoting the number of content units in a given device cache. 
\begin{algorithm}\footnotesize
\caption{Energy Prioritized D2D Caching Algorithm (EPDC)}\label{alg:EPDC}
\begin{algorithmic}
\STATE{}
\STATE{\textbf{EPDC(}$\boldsymbol{C,u', \HUDeviceCacheSize}$)\{}
\STATE{}
\STATE{$Capacity$ $\gets$ $TotalSize(C)$;}
\IF{($Capacity+\size{u'}\leq \HUDeviceCacheSize$)}
    \STATE{return$~C\cup\{u'\};$ \%the set kept in the cache}
\ELSE
        \STATE{\%sort units in $C$ based on their $E_{all}^(i)$ values (in~(\ref{eq:profit_all})) in descending order}
        \STATE{$\%C_{sort}=\{c_1,c_2,...,c_k\}$ \text{with $c_1$ having largest $E_{all}$, $c_k$ lowest}}
        \STATE{$C_{sort}$ $\gets$ sort($C$,~$E_{all}$,~DESCEND);}
        \STATE{j = k;}
        \WHILE{($j\geq1$)}
        \STATE{$C_{sort} \gets C_{sort} \setminus  \{c_j,c_{j+1},...c_k\};$}
        \IF{($Capacity+\size{u'}-\sum_{\theta=j}^{k}\size{c_{\theta}} \leq \HUDeviceCacheSize$)}
            \STATE{return $C_{sort}\cup\{\mathbf{u'}\};$ \%the set decided to be kept in the cache}
        \ENDIF
        \STATE{$j \gets j-1$;}
        \ENDWHILE
\ENDIF
\STATE{\}}
\end{algorithmic}
\end{algorithm}

\subsection{Time complexity analysis}
Based on our complexity analysis, $OPT$ algorithm has infeasible operation time. As shown in Algorithm~\ref{alg:dyn}, for all cached units $OPT$ has an iteration range by the cache capacity with $\delta$ incremental units. For devices we take $\deltaDev = 0.01$ Mbs and for the BS cache $\deltaBS = 0.1$ Mbs respectively. The average time consumed for devices and the BS cache by all techniques is provided in Table~\ref{table:Complexity}. Note that we test the time complexity in a laptop device with Intel core i7-6500 CPU @2.5GHz and 8 GB RAM in MATLAB R2019 environment.  
We apply any chosen algorithm not only in devices but also in the BS cache for the sake of completeness and hence due to the large BS cache capacity even in a setup with moderate number of contents, we present the impracticality of applying $OPT$ in real-time or near real-time scenarios. With $\HUDeviceCacheSize$ = 150 Mbs and $\BSCacheSize$ = 2.8 Gbs even with 150 contents, $OPT$ performs poorer than $EPDC$ in terms of service time especially for the BS caching. 
This is because $EPDC$ algorithm is only bounded by the sorting of content units but not cache capacity while $OPT$ algorithm acts an iteration over the cache capacity for each in-cache content unit. 
\begin{table}[ht]
	\centering
	\caption{Time complexity.}
		\begin{tabular}{ | l | p{3cm} | p{3cm} |}
			\hline
			\textbf{Technique} & \textbf{Total Local Caching Time (s)} & \textbf{Total BS Caching Time (s)} \\ \hline
			$LRU$       & 0.57  & 1.11       \\ \hline
			$\PDC$      & 0.53  & 2.65       \\ \hline  
			$SXO$  & 0.82  &  9.69          \\ \hline
			$OPT$       & 18.94 & 7319.11   \\ \hline
			$EPDC$      & 6.01  & 5.80       \\ \hline
		\end{tabular}
	\label{table:Complexity}
\end{table}

\section{Performance Metrics}\label{sec:performanceMetrics}
We analyze optimal scalable video caching algorithm ($OPT$) in D2D cellular networks with backhaul support and energy prioritized D2D caching ($EPDC$) algorithm in terms of \textit{service rate} and \textit{energy expenditure}. The system parameters are listed in Table~\ref{table:sysParameters}. We compare $OPT$ and $EPDC$ algorithms to the commonly used baseline technique Least Recently Used ($LRU$). $LRU$ technique does not consider content features.
Therefore, it used to compare the improvement of our algorithms over a classical non-content aware cache replacement technique. In the literature, popularity based caching policies are also broadly utilized~\cite{6761239, 7841535, 7841508}. As another comparison technique, we utilize the Popularity-driven caching ($\PDC$) algorithm~\cite{SINEMKAFILOGLU2020107083}, 
that manages the cache replacement with the aim of keeping popular contents in cache with a greater \textit{probability}. 
This technique is aware of content attribute popularity and decides on the evictions accordingly. As multimedia content caching is the main focus of our algorithms, we compare our algorithms to this content-aware cache replacement technique $\PDC$. We also compare our algorithms to the $SXO$ technique~\cite{1542018}. In this technique, content features size and order (access rate) are considered for the cache replacement decision where large contents with low access rate are evicted first. Note that we will investigate the service rate and energy expenditures in different operation modes. In that regard, the energy consumption of our system consists of the following parts: 
\begin{itemize}
    \item $E_{loc}$ for local hits 
    \item $E_{D2D}$ for D2D transmission 
    \item $E_{BS}$ for direct services of the BS 
    \item $E_{BS(U)}$ for indirect services of the BS 
    \item $E_{block}$ due to blocked services 
\end{itemize}
$E_{loc}$, $E_{D2D}$ and $E_{block}$ are defined as in our previous work~\cite{PIMRC}. 
\begin{table}[ht]
\centering
	\caption{System parameters.}
	\begin{tabular}{ | p{1.4cm} | p{6.2 cm}| }
			\hline
			\textbf{Parameter} & \textbf{Explanation}  \\ \hline
			$\BSPower$ & The transmission power consumption of the BS  \\ \hline
			$\BSUnivPower$ & The reception power consumption of the BS  \\ \hline
			$C_{BS}(n)$ & The channel capacity between the BS and the $n^{th}$ device  \\ \hline
			$C_{BS(U)}$ & The capacity between the universal content repository and the BS  \\ \hline
			$N$ & The total number of devices  \\ \hline
			$U$ & The set of content units uniquely identifiable by content, chunk, layer id  \\ \hline
			$r_u$ & The request for the content unit $u$  \\ \hline
			$\size{u}$ & The size of the content unit $u$  \\ \hline
			$S^{BS}_{(n)}$ & The set of services from the BS to  $n^{th}$ device \\ \hline
			$S^{BS(U)}_{(n)}$ & The set of services from the universal content repository to the $n^{th}$ device across the BS \\ \hline
	\end{tabular}
	\label{table:sysParameters}
\end{table}	
One of our new contributions is the BS integration to our D2D network architecture. Therefore, we define the BS energy expenditure branched into \textit{i) direct services} and \textit{ii) indirect services} as $E_{BS}$ and $E_{BS(U)}$ in~(\ref{eq:E_BS}) and~(\ref{eq:E_BSU}), respectively. 
$E_{BS}$ is the overall transmission energy consumption of the BS serving requests directly from the BS cache whereas $E_{BS(U)}$ is the energy consumption for services from the universal source across the BS to requesters. 
\begin{align}
    E_{BS} &:=  \sum_{u \in U} \sum_{\substack{n \in N}} \sum_{r_u \in S^{BS}_{(n)}} P_{BS}\cdot\frac{|\size{u}|}{C_{BS}(n)}\label{eq:E_BS}\\
    E_{BS(U)} &:= \sum_{u \in U} \sum_{\substack{n \in N}} \sum_{r_u \in S^{BS(U)}_{(n)}} \frac{P_{BS}\cdot|\size{u}|}{C_{BS}(n)}+\frac{P_{BS(U)}\cdot|\size{u}|}{C_{BS(U)}}\label{eq:E_BSU}
\end{align}

By the summation of all the expenditures, we obtain  $E_{total}:=E_{loc}+E_{D2D}+E_{BS}+E_{BS(U)}+E_{block}$ the total energy consumption of our system.

\section{Performance Evaluation}\label{sec:results}

We perform simulations for varying system parameters \textit{cache capacity} and \textit{D2D transmission radius}, and investigate how these parameters impact the network for the \textit{service rate} and \textit{energy expenditure}. We compare $EPDC$ and $OPT$ algorithms to the $LRU$, $\PDC$~\cite{SINEMKAFILOGLU2020107083} and $SXO$~\cite{1542018}. The simulation parameters are listed in Table~\ref{table:simParameters}. In this list, the content unit popularity and request parameters ($s$, $\lambda$, $k$, $p_k$), the power of system components (local, BS, device), local hit and BS reception power tune parameters ($\theta_{loc}$, $\theta_{BS}$), cache capacities ($\HUDeviceCacheSize$, $\BSCacheSize$), channel parameters, device density $\lambdaNHU$ and the maximum transmission radius of D2D mode $\RDTwoD$ are defined. In Table~\ref{table:simParameters}, the local and D2D power consumptions are in mW scale while the BS transmission/reception powers are at the order of Watts as observed in practical systems. 
The channel parameters $B$, $d_0$ and $n_{D2D}$ are adopted from our previous study~\cite{PIMRC}. We take the path loss exponent of the BS transmissions $n_{BS}$ greater than that of the D2D mode $n_{D2D}$ as broadly employed in the literature~\cite{8758385,7056528}. We have an event-based simulator that processes content unit request arrivals and service completions. Since we elaborate on the caching of multimedia contents, we simulate content requests according to a popularity scheme that utilizes Zipf distribution with parameter $s$. Each content is partitioned into chunks and the requests for these chunks follow the Weibull distribution with parameters $\lambda$ and $k$. Due to the scalable content structure, our contents are layered into base and enhancement. The high quality (HQ) content consumption with both layers is realized by the HQ request probability $p_{k}$. For the event-based simulator that acts upon each content unit request arrival/completion, network operations require power and channel parameters to actualize the arrivals. In that regard, these parameters are provided in the Table~\ref{table:simParameters}. In the following subsections, we vary the device cache capacity $\HUDeviceCacheSize$ and D2D transmission radius $\RDTwoD$ to elaborate on how the D2D cache and service capability impact the system.

\begin{table}[ht]
	\centering
	\caption{Simulation parameters.}
		\begin{tabular}{ | l | p{4.9 cm}| p{1.5 cm} | }
			\hline
			\textbf{Symbol} & \textbf{Explanation} & \textbf{Value}  \\ \hline
			$T_{sim}$ & The simulation duration & 400 s \\ \hline	
			$s$ & The Zipf distribution skewness parameter & 1\\ \hline
			$\lambda$ & The Weibull distribution scale parameter & 5 \\ \hline
			$k$ & The Weibull distribution shape parameter & 0.8\\ \hline
			$p_{k}$ & The probability of high quality contents being requested & 1 \\ \hline
		
			$\localPower$ & The power consumption of local content unit retrieval & 40 mW \\ \hline
			$\DTwoDPower$ & The transmission power consumption of a device & 80 mW \\ \hline
			$\BSPower$ & The transmission power consumption of the BS  & 6 W \\ \hline
			$\BSUnivPower$ & The reception power consumption of the BS  & 1.2 W \\ \hline
			$\theta_{loc}$ & The local hit service power parameter & 2\\ \hline
			$\theta_{BS}$ & The base station reception power parameter & 5\\ \hline
			$\HUDeviceCacheSize$ & The device cache capacity & 150 Mbits \\ \hline
			$\BSCacheSize$ & The base station cache capacity & 2.8 Gbits \\ \hline
			$B$ & The channel bandwidth & 2 MHz \\ \hline
			$N_0$ & The noise power density &  -158 dBm \\ \hline
			$d_0$ & The reference distance of device antenna & 1 m \\ \hline
			$n_{D2D}$ & The path loss exponent of D2D transmission & 3 \\ \hline
			$n_{BS}$ & The path loss exponent of BS transmission & 4.2 \\ \hline
			$\lambdaNHU$ & The mean device density across the cell according to Point Poisson Process (PPP) & 0.0015 $\frac{user}{m^2}$ \\ \hline
		
			$\RDTwoD$ & The maximum D2D transmission operation radius
			& 200 m\\ \hline
		\end{tabular}
	\label{table:simParameters}
\end{table}

\subsection{Impact of device cache capacity $\HUDeviceCacheSize$}
In this section, we investigate a key component in our system, the device cache capacity $\HUDeviceCacheSize$. We enlarge/shrink the cache capacity $\HUDeviceCacheSize$ to observe how it impacts the caching techniques. 

\begin{figure}[h]
\centering
\subfloat[\normalsize{Locally served content units (B: of base layer,\\ E: of enhancement layer, S: successful).}]{\includegraphics[width=0.9\columnwidth]{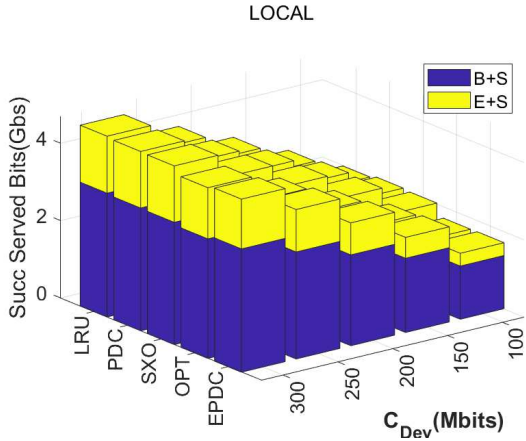}
\label{fig:C_bits_local}} \\
\subfloat[\normalsize{The energy consumed for the content units (B: of base layer, E: of enhancement layer, S: successful, F: fail, \textit{Remark:  the energy consumption of failed services is very small (green and yellow components) compared to the energy of successful services}).}]{\includegraphics[width=0.9\columnwidth]{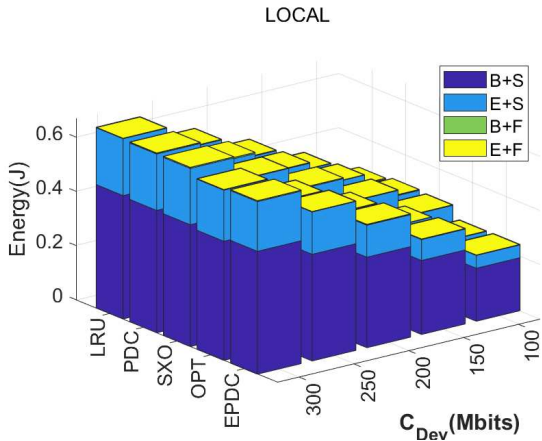}
\label{fig:C_energy_local}}
\caption{Results for the local caches.}
\label{fig:C_local}
\end{figure}

First, we compare our caching technique $EPDC$ with the $OPT$ and the alternative strategies $LRU$, $\PDC$ and $SXO$ in terms of the local hit service rate and energy consumption. 
For any fixed $\HUDeviceCacheSize$ value, the energy optimizing $OPT$ attains the smallest local service rate and in return the lowest local energy expenditure given in Fig.~\ref{fig:C_bits_local} and~\ref{fig:C_energy_local}, respectively. 
Compared to the optimal case, our heuristic $EPDC$ is slightly worse in terms of energy consumption for local hits. However, the energy improvement over  $LRU$, $\PDC$ and $SXO$ manifests the utilitarian nature of our proposal for local hits. 
This improvement of $EPDC$ over the alternatives becomes greater for decreased cache capacity. In the large cache capacity regime, the cache capacity has a greater impact on the energy expenditure as requested contents can be found locally with a greater probability. With decreasing cache capacity, the impact of cache replacement procedure can be observed more clearly.
For $\HUDeviceCacheSize= $ 300 Mbits the local hit energy consumption improvement rate of $EPDC$ over $LRU$ ($\PDC$ / $SXO$) is 4.0\% from 0.67 to 0.64 J (3.5\% from 0.66 to 0.64 J / 3.2\% from 0.66 to 0.64 J). With decreasing $\HUDeviceCacheSize$ to 100 Mbits, this improvement over $LRU$ ($\PDC$ / $SXO$) is amplified to 14.4\% from 0.29 to 0.24 J (17.0\% from 0.29 to 0.24 J / 15.4\% from 0.29 to 0.24 J).

In all caching techniques, for increasing $\HUDeviceCacheSize$, the direct BS service usage and energy expense decline since large device caches entail increased local and D2D service utilization. 
For an arbitrary $\HUDeviceCacheSize$ value, our $EPDC$ technique achieves less direct BS mode energy consumption than $LRU$, $\PDC$ and $SXO$ techniques. The direct BS energy improvement rate of $EPDC$ over $LRU$ ($\PDC$) is at its best at $\HUDeviceCacheSize =$ 200 Mbits with 7.3\% from 1.65 to 1.53 kJ (12.5\% from 1.75 to 1.53 kJ). This improvement of $EPDC$ over $SXO$ reaches to its maximum at $\HUDeviceCacheSize=$ 300 Mbits with 12.4\% 0.99 to 0.87 kJ. $EPDC$ falls back from the $OPT$ for the BS energy consumption. However, this degradation is at most 6.0\% for $\HUDeviceCacheSize =$ 100 Mbits. Nevertheless, the evident improvement over $LRU$, $\PDC$ and $SXO$ demonstrates the advantage of our heuristic $EPDC$ for the direct BS energy consumption.

\begin{figure}[ht]
	\centering
	\includegraphics[width=0.9\columnwidth]{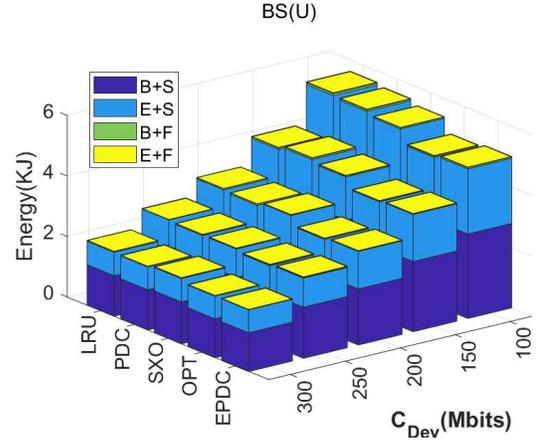}
	\caption{The energy consumed for retrievals in BS(U) mode (B: of base layer, E: of enhancement layer, S: successful, F: fail).}
	\label{fig:C_energy_BS_U}	
\end{figure}
In our construct for indirect services of the BS, a similar reasoning with direct BS services and corresponding energy expenditures is valid. $EPDC$ consumes less energy than $LRU$, $\PDC$ and $SXO$ shown in Fig.~\ref{fig:C_energy_BS_U} for all $\HUDeviceCacheSize$ values.

The indirect BS energy expenditure improvement rate of $EPDC$ over $LRU$ ($\PDC$ / $SXO$) is at most 11.1\% from 5.63 to 5.00 kJ (at most 10.8\% from 4.38 to 3.91 kJ / 8.1\% from 3.40 to 3.12 kJ) for $\HUDeviceCacheSize$ = 100 (150 / 200) Mbits. 
$EPDC$ achieves indirect BS energy results close to the optimal case with at most 1.6\% difference. Hence, our heuristic is practical for the energy reduction of the BS regarding the indirect BS services.

\begin{figure}[ht]
	\centering
	\includegraphics[width=0.9\columnwidth]{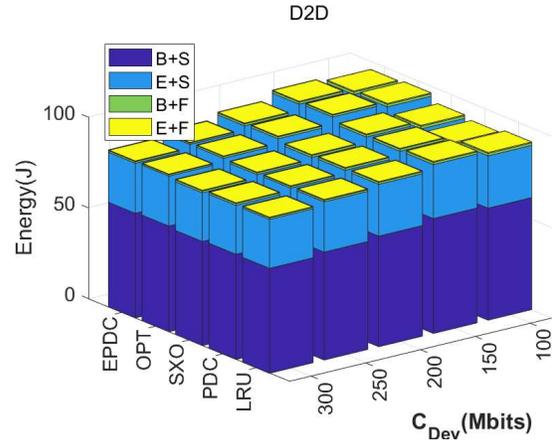}
	\caption{The energy consumed for retrievals in D2D mode (B: of base layer, E: of enhancement layer, S: successful, F: fail).}
	\label{fig:C_energy_D2D}	
\end{figure}
For the D2D analysis, the corresponding service rate is not affected by the varying $\HUDeviceCacheSize$ values evidently for different caching policies. In correlation, the energy consumption of D2D usages are not significantly impacted by the varying device cache capacity $\HUDeviceCacheSize$ for any given caching technique. The reason is that the varying cache capacity particularly impacts the local hit results. When we compare $EPDC$ to the alternatives $LRU$, $\PDC$ and $SXO$ for any $\HUDeviceCacheSize$, we observe a slight increase in D2D service rate. In that regard, the energy consumption due to D2D services for $EPDC$ is increased very slightly compared to $LRU$, $\PDC$ and $SXO$ strategies. 
The greatest increase in D2D energy consumption of $EPDC$ over $LRU$ ($\PDC$ / $SXO$) is 3.6\% from 92.43 to 95.78 J (5.6\% from 90.49 to 95.78 J / 4.0\% from 92.12 to 95.78 J) for $\HUDeviceCacheSize =$ 100 Mbits as shown in Fig.~\ref{fig:C_energy_D2D}.
The alternatives do not consider prospective energy consumption of the local, D2D, BS and BS(U) modes. However, our technique makes use of them and favors the D2D paradigm in contrast to the BS transmissions since D2D technique is more beneficial than cellular (BS) content transmissions in terms of energy cost. Thereof, we observe slight increase in D2D service rate and corresponding energy consumption in $EPDC$ technique compared to the alternative counterparts. We expect to observe a major increase in D2D usage due to its low power consumption. However, the limited device capacities compared to the BS and the high power of the BS than devices are dominating factors against reducing the overall network energy consumption of $EPDC$. Thereof, we do not observe a significant change in D2D preference as in BS preference between different policies.

\begin{figure}[ht]
	\centering
	\includegraphics[width=0.9\columnwidth]{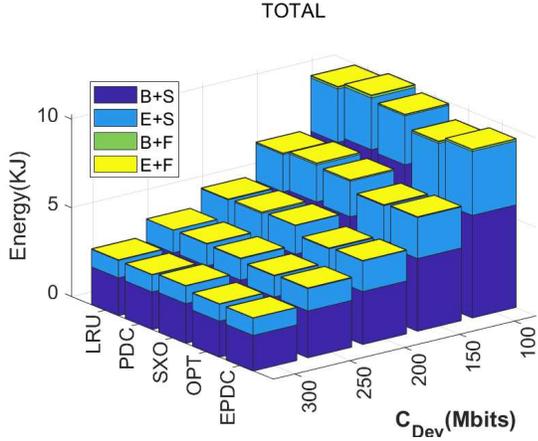}
	\caption{The total energy consumed for retrievals and local hits (B: of base layer, E: of enhancement layer, S: successful, F: fail).}
	\label{fig:C_energy_ALL}
\end{figure}
Finally, we focus on the total service rate and energy consumption of the network.
As already mentioned, the D2D service rate is not impacted by the varying $\HUDeviceCacheSize$ for different caching policies. However, for increasing $\HUDeviceCacheSize$, the local service capacity rises (in Fig.~\ref{fig:C_bits_local}) while the BS service capacity reduces in all caching policies. The increase in local and decrease in BS service capacities balance each other and the total service capacity 
is not changed evidently for varying $\HUDeviceCacheSize$ in each caching policy. 
In all policies, with increasing $\HUDeviceCacheSize$ the energy consumption trends for local service/D2D/BS/BS(U) service types are consistent with the corresponding service capacity behaviour. 
Although for all caching strategies the total service capacity of the network is not affected by different $\HUDeviceCacheSize$ values, with increasing $\HUDeviceCacheSize$ the total network energy consumption declines for all caching strategies and hence the network-wide energy efficiency is improved.

As shown in Fig.~\ref{fig:C_energy_ALL}, our proposed algorithm $EPDC$ falls back from the optimal with at most 3.5\% (for $\HUDeviceCacheSize =$ 100 Mbits) in our investigated $\HUDeviceCacheSize$ range and $EPDC$ has very close results to the optimal policy especially for large $\HUDeviceCacheSize$'s.
Although for fixed $\HUDeviceCacheSize$ values the measured total network service capacity does not change significantly between different techniques, the observed total network energy expenditure of our technique $EPDC$ is less and thus energy-wise more rewarding compared to $LRU$, $\PDC$ and $SXO$ 
as depicted in Fig.~\ref{fig:C_energy_ALL}. The greatest total network energy improvement of $EPDC$ compared to $LRU$ ($SXO$) is with 6.7\% from 5.08 to 4.74 kJ (8.4\% from 5.18 to 4.74 kJ) for $\HUDeviceCacheSize =$ 200 Mbits. $EPDC$ attains the maximum improvement over $\PDC$ at $\HUDeviceCacheSize =$ 150 Mbits with 9.6\% from 7.15 to 6.46 kJ. 
Thus, we observe the beneficial nature of our proposal for energy-sensitive network scenarios.

\subsection{Impact of D2D transmission radius $\RDTwoD$}
We investigate how different settings of D2D transmission impact the system. In that regard, we vary the D2D transmission radius $\RDTwoD$. 
\begin{figure}
\centering
\subfloat[Locally served content units (B: of base layer, E:\\ of enhancement layer, S: successful).]{\includegraphics[width=0.9\columnwidth]{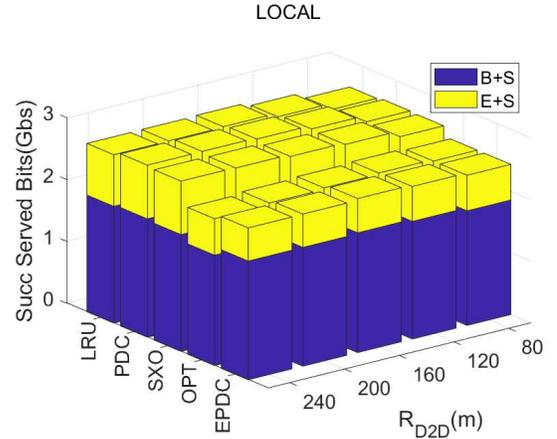}
\label{fig:R_D2D_bits_local}}\\ 
\subfloat[The energy consumed for the content units (B: of base layer, E: of enhancement layer, S: successful, F: fail).]{\includegraphics[width=0.9\columnwidth]{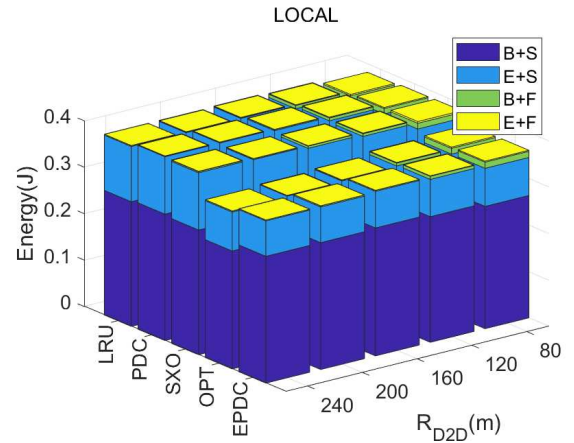}
\label{fig:R_D2D_energy_local}}
\caption{Results for the local caches.}
\label{fig:R_D2D_local}
\end{figure}
Initially, we study the local hit service capacity and corresponding energy consumption. 
As depicted in Fig.~\ref{fig:R_D2D_bits_local}, the local hit service capacity changes negligibly with varying $\RDTwoD$ for all caching strategies. When a content is available in the local cache, the service is locally provided and other network resources are not used. Hence, the D2D features do not impact the local hit service capacity. Besides, for all caching policies the local hit energy consumption is not affected by $\RDTwoD$ (Fig.~\ref{fig:R_D2D_energy_local}).  

When we compare our proposal $EPDC$ to $LRU$, $\PDC$, $SXO$ and $OPT$, $EPDC$ achieves lower local service capacity and, in correspondence with this, local energy consumption improvement compared to $LRU$, $\PDC$ and $SXO$ caching policies for any fixed $\RDTwoD$ value in the inspection range. Given in Fig.~\ref{fig:R_D2D_energy_local}, the greatest improvement for local hit energy consumption of our proposal over the alternative $LRU$ ($\PDC$ / $SXO$) occurs at $\RDTwoD = $ 240 m attaining 9.8\% from 0.39 to 0.35 J (11.6\% from 0.40 to 0.35 J / 11.1\% from 0.39 to 0.35 J). 
Even the least local energy consumption improvement of $EPDC$ over $LRU$ ($\PDC$ / $SXO$) is 5.6\% (6.2\% / 6.3\%) at $\RDTwoD = $ 80 m.  
As expected, $EPDC$ is not as good as the optimal caching in terms of local energy expenditure. Nevertheless, the largest rise observed in terms of local energy consumption of our proposal over the optimal policy is 3.3\% only (from 0.35 to 0.34 J) at $\RDTwoD$ = 240 m. Hence, the local energy expenditure improvement of $EPDC$ over the baseline $LRU$ and other comparison strategies is an indicator of the energy-wise utility of $EPDC$ for local services.

\begin{figure}[ht]
	\centering
	\includegraphics[width=0.9\columnwidth]{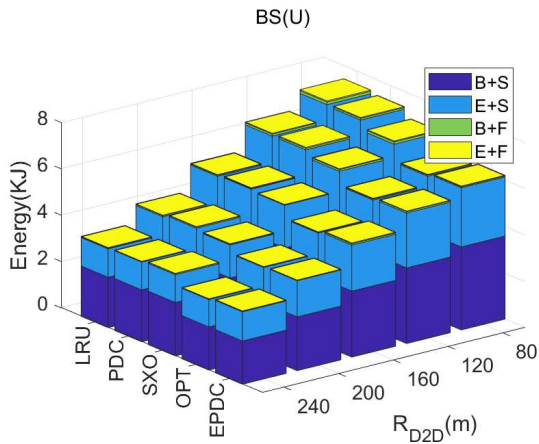}
	\caption{The energy consumed for retrievals in BS(U) mode (B: of base layer, E: of enhancement layer, S: successful, F: fail).~~}
	\label{fig:R_D2D_energy_BS_U}
\end{figure}

Next, we perform an analysis on the direct BS services. The increasing D2D transmission radius $\RDTwoD$ amplifies the chance of finding requested content units in neighbours and subsequently the D2D service allocations are boosted. The BS services are less required and the direct BS service rate declines for all caching policies. In accordance with this change, in all policies the direct BS mode energy consumption reduces with increasing $\RDTwoD$. In our analysis, similarly a decrease is observed for indirect BS service rates with respect to increasing $\RDTwoD$. Also, decrease in indirect BS energy consumption with respect to increasing $\RDTwoD$'s is observed at all caching techniques as shown in Fig.~\ref{fig:R_D2D_energy_BS_U}. 

For $EPDC$, the services supplied from the universal source across the BS (indirect BS services) hits lower service rate than the other algorithms for all $\RDTwoD$'s. Accordingly, as shown in Fig.~\ref{fig:R_D2D_energy_BS_U}, $EPDC$ achieves improvement in indirect BS energy consumption over the comparison algorithms for all $\RDTwoD$'s. The optimal policy and our proposed algorithm $EPDC$ are designed for minimizing the total network energy expenditures. Hence, omitting such long routes (from the universal source across the BS to requesters) with large power levels is instrumental for reducing overall system energy consumption. 
Considering policy comparisons for direct BS services, the corresponding energy expenditure improvement over $LRU$ ($\PDC$ / $SXO$) is 12.2\% (19.4\% / 10.2\%) for the largest $\RDTwoD =$ 240 m. $EPDC$ falls back to the $LRU$, $\PDC$ and $SXO$ with decreasing $\RDTwoD$ in terms of direct BS energy consumption because the D2D service rate declines with decreasing $\RDTwoD$ and this enforces the system to get direct BS services for units available in the BS cache.

\begin{figure}[ht]
	\centering
	\includegraphics[width=0.9\columnwidth]{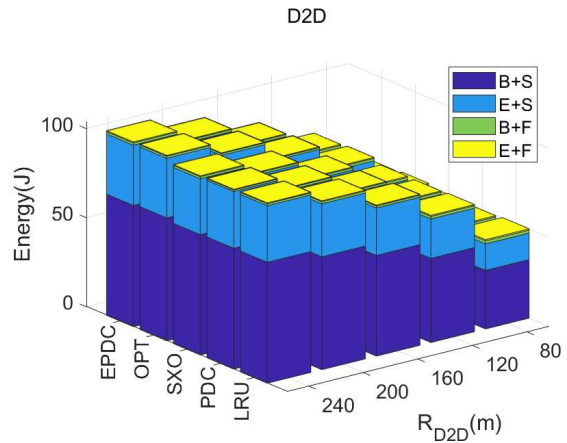}
	\caption{The energy consumed for retrievals in D2D mode (B: of base layer, E: of enhancement layer, S: successful, F: fail).~~~~~~~~~~~~~~~~~}
	\label{fig:R_D2D_energy_D2D}
\end{figure}
For all caching policies, the D2D service rate and corresponding energy consumption increases with increasing $\RDTwoD$ as illustrated in Fig.~\ref{fig:R_D2D_energy_D2D}. The reason is that the number of devices available in reception range of requesters increases in that case. Consequently, the probability of finding requested content unit(s) in a neighbour device is higher according to~(\ref{eq:D2Dprob}) and hence D2D service rate is improved. 
Our algorithm $EPDC$ achieves slightly better rate and correspondingly, slightly larger energy consumption compared to alternative algorithms as shown in Fig.~\ref{fig:R_D2D_energy_D2D}. 
As already explained in the previous subsection, the high power level of the BS is the key factor in determining the overall network energy consumption and $EPDC$ algorithm chokes the BS selection rather than increasing the D2D mode selection.
Hence, no significant D2D usage change is observed between different policies.

\begin{figure}[ht]
	\centering
	\includegraphics[width=0.9\columnwidth]{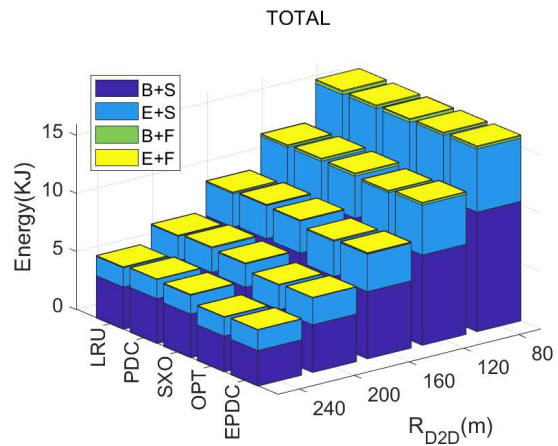}
	\caption{The total energy consumed for retrievals and local hits (B: of base layer, E: of enhancement layer, S: successful, F: fail).}
	\label{fig:R_D2D_energy_ALL}
\end{figure}

We also inspect the total service rate and energy expenditure of the network with respect to varying D2D transmission radius $\RDTwoD$. 
For all policies, the total network service rate rises with increasing $\RDTwoD$ from 80 m to 120 m. With further increase, it becomes saturated in all policies.
For all caching algorithms, we observe an increase in the total network service rate for the greatest D2D service improvement rate ($\RDTwoD$ from 80 m to 120 m) that is not canceled out by the BS service rate drop. This is because the D2D service is more crucial with its larger rate compared to the other service factors in determining total network service rate. 
With further increasing $\RDTwoD$ values, we again observe recession in BS service rates and this time they cancel the increase in D2D service rates. This is due to the slower D2D service improvement. 
As already mentioned, in all caching techniques the BS energy expenditure decreases, the D2D energy consumption rises whereas the local expenditure is stagnant with increasing $\RDTwoD$'s. The BS services take longer service durations with higher power levels in contrast to D2D mechanism. Thus, the BS manifests itself with its great energy consumption impact. Overall, for all caching strategies in Fig.~\ref{fig:R_D2D_energy_ALL}, the total energy expense of the network decreases with increasing $\RDTwoD$'s.

For fixed $\RDTwoD$ values, the total network service rate does not change evidently among different caching techniques. However, as shown in Fig.~\ref{fig:R_D2D_energy_ALL} for all $\RDTwoD$'s, the overall energy consumption of the network by $EPDC$ is less than other algorithms. This energy expenditure improvement of $EPDC$ rises with increasing $\RDTwoD$. This is basically due to the difference in BS usage and corresponding energy consumption shown in Fig.~\ref{fig:R_D2D_energy_BS_U}. 
The greatest decrease in the total network energy consumption of $EPDC$ compared to $LRU$ ($\PDC$ / $SXO$) is for $\RDTwoD =$ 240 m with 9.5\% from  5.38  to 4.87 kJ (12.4\% from 5.56 to 4.87 kJ / 10.6\% 5.45 to 4.87 kJ). 
When we compare our proposal $EPDC$ with the optimal counterpart, it slightly performs worse with at most 2.4\% performance degradation in the total network energy consumption at $\RDTwoD =$ 240 m. In spite of this fall back, we observe improvement of $EPDC$ over the compared algorithms for $\RDTwoD\geq 120$ m. As already mentioned, no evident change in total network service rate is observed between different caching techniques for fixed $\RDTwoD$ values. Hence, especially for large $\RDTwoD$'s, $EPDC$ algorithm becomes more energy efficient than the alternative ones.

\section{Conclusions}\label{sec:conclusion}
D2D networking is an attractive technique for improving system energy efficiency. Similarly, caching is a beneficial scheme used for boosting system capacity and reducing energy consumption in wireless networks. To this end, we study the cache replacement management in cellular D2D networks in this work. We formulate and solve the optimal caching problem minimizing the energy consumption in cellular D2D network for the video content consumption scenario. Besides, we propose an energy prioritized D2D caching algorithm considering the prospective energy consumption with the aim of reducing the overall system energy burden. 
For rigorous system analysis, we inspect the service rate and energy consumption of different service modes with respect to varying device cache capacity and D2D transmission range. Based on our evaluations, our proposed algorithm EPDC provides network-wide energy efficiency gains compared to alternative
algorithms, especially for larger D2D transmission ranges.

\section*{Acknowledgments}
This work was supported by the Scientific and Technical Research Council of Turkey (TUBITAK) under grant number 116E245.


\bibliographystyle{IEEEtran}
\bibliography{references}

\appendix
\section{Utility Functions}
For the calculation of the prospective overall expected energy consumption $\profitAll{u}$ of any \textit{content unit} $u$ that is mapped to some \textit{content unit tuple} $\{i,j,k\}$, we utilize content unit availability probabilities. 
The availability probability in the local cache of a requester of some arbitrary content unit $u$ (with the one-to-one and onto correspondent \textit{content unit tuple} $\{i,j,k\}$) is defined in~(\ref{eq:localprob}) as $\weightLocal{u}$. The multiplication of the $i^{th}$ content request probability $p_i$, the $j^{th}$ chunk request probability $p_j$ and the $k^{th}$ layer request probability $p_k$ gives request probability of \textit{content unit tuple} $\{i,j,k\}$ (\textit{content unit} $u$). Aside from the request characteristic, the requester device 
storage capability is eminently prominent for the calculation of the content unit availability probability at a requester device. For the integration of this aspect, the device capacity $\HUDeviceCacheSize$ is normalized over the total size of all content space ($\sum_{all}\size{u}$) by $\fitfunctionDevice$ function in~(\ref{eq:fitD2D}).
\begin{align}\label{eq:localprob}
\weightLocal{u} &:=  \probI \cdot \probJ \cdot \probK \cdot \fitfunctionDevice\\
\label{eq:fitD2D}
\fitfunctionDevice  &:= \frac{\HUDeviceCacheSize}{\sum_{all} \size{u}}
\end{align}
The availability probability of some arbitrary content unit $u$ in the BS cache $\weightBS{u}$ in~(\ref{eq:BSprob}) is calculated similarly. The corresponding normalization function $\fitfunctionBS$ that outputs the BS storage capability is given in~(\ref{eq:fitBS}).

\begin{align}\label{eq:BSprob}
\weightBS{u}       &:= \probI \cdot \probJ \cdot \probK \cdot \fitfunctionBS \\
\label{eq:fitBS}
\fitfunctionBS      &:= \frac{\BSCacheSize}{\sum_{all} \size{u}}
\end{align}

$\Nneighbours$ is defined as the total number of devices that are at most $R_{D2D}$ away from a requester. 
The availability probability of some content unit $u$ in at least one of the neighbour devices of a requester $\weightDTwoD{u}$ is defined in~(\ref{eq:D2Dprob}). $(1-\weightLocal{u})^{\Nneighbours}$ gives the probability of a requested content unit $u$ not being available in any neighbouring device of a requester. By the complement, we obtain the availability probability of some content unit $u$ being in at least one of the neighbours of a requester device.  
\begin{align}\label{eq:D2Dprob}
\weightDTwoD{u} &:= 1-  (1-\weightLocal{u})^{\Nneighbours}
\end{align}

\end{document}